\documentclass[
reprint,
nofootinbib,
amsmath,amssymb,
aps,
floatfix 
]{revtex4-2}

\usepackage{graphicx}
\usepackage{dcolumn}
\usepackage{bm}
\usepackage[colorinlistoftodos]{todonotes}
\usepackage[export]{adjustbox}%
\usepackage{natbib}
\usepackage{lineno}
\usepackage{amsthm}

\usepackage{xcolor}
\colorlet{rcolor}{black}  

\renewcommand{\Pr}[1]{\ensuremath{\mathsf{Pr}\!\left\{#1\right\}}}

\DeclareFontFamily{U}{stix2bb}{}
\DeclareFontShape{U}{stix2bb}{m}{n} {<-> stix2-mathbb}{}

\begin{document}
	
	\title{How competition propels scientific risk-taking}
	
	\author{Kevin Gross}
	\email{krgross@ncsu.edu}
	\affiliation{Department of Statistics \\ North Carolina State University \\ Raleigh, NC USA}

	\author{Carl T. Bergstrom}
	\email{cbergst@u.washington.edu}
	\affiliation{Department of Biology \\ University of Washington \\ Seattle, WA USA \\ }

	\date{\today}
	
	\begin{abstract}
		In science as elsewhere, attention is a limited resource and scientists compete with one another to produce the most exciting, novel and impactful results. We develop a game-theoretic model to explore how such competition influences the degree of risk that scientists are willing to embrace in their research endeavors. We find that competition for scarce resources---for example, publications in elite journals, prestigious prizes, and faculty jobs---motivates scientific risk-taking and may be important in counterbalancing other incentives that favor cautious, incremental science. Even small amounts of competition induce substantial risk-taking. Moreover, we find that in an ``opt-in'' contest, increasing the stakes induces increased participation---which crowds the contest and further impels entrants to pursue higher-risk, higher-return investigations. The model also illuminates a source of tension in academic training and collaboration. Researchers at different career stages differ in their need to amass accomplishments that distinguish them from their peers, and therefore may not agree on what degree of risk to accept.
		
		\end{abstract}
	\maketitle
\section{Introduction}

The practice of science can be a competitive business that often pits researchers against one another in contests of various forms \cite{hull2010science}. Scientists battle to obtain grant funding \cite{gross_contest_2019,azoulay2020scientific}, compete to hire the most talented or best-credentialed scholars into their departments \cite{wapman2022quantifying}, and race against one another to establish priority for their discoveries \cite{merton_priorities_1957}.  Even when working on separate problems, scientists compete with one another to produce work that stands out for its novelty, excitement, and impact. These efforts play out as competitions for publications in high-profile journals,  for prizes, and perhaps most importantly, for faculty jobs. Competition in all of these domains appears to have steadily intensified over the past few decades. Tenure-track positions, meanwhile, have become increasingly scarce and require increasingly impressive credentials to land \cite{cyranoski2011education}. Moreover, outcomes in each domain are connected to outcomes in the others \cite{waaijer2018competition}. For example, high-profile publications play a more important role in career success than ever before \cite[e.g.,][]{heckman2020publishing,verma_impact_2015}. 

We argue here that the contest for exciting results---much like the  priority rule and the race for first publication \cite{arrow1962economic,dasgupta1987information,dasgupta1987simple,strevens2003,kleinberg2011mechanisms,bergstrom2016scientists,zollman2018credit,heesen2018reward}---structures the research strategies of scientists and influences the allocation of effort across research problems. We focus on the notion of risk. In science, high risk is often associated with high return \cite{rzhetsky2015choosing,stephan2015economics}, yet a host of structural incentives often drive investigators to pursue more conservative strategies, \cite{foster2015tradition, franzoni2022funding,gross2021ex,gross2024rationalizing}.
Scientific risk is a complex concept \cite{franzoni2023uncertainty}, but here we use a simple definition: a high-risk, high return project is one that has a high probability of failing but which would have major impact should it succeed.  When scientists choose problems, they have to account for both the chances of failure and the rewards of success.

In this paper we explore how the competitive nature of scientific contests affects the balance that researchers strike between risk and reward when designing their individual research programs, and how this in turn shapes the aggregate landscape of scientific activity.  To do so, we develop a game-theoretic model in which investigators select projects while competing with one another for scientific impact and its attendant professional rewards.  We present two variations of the model. In the first, researchers are obligated to participate in the contest, as might be the case with junior researchers competing for faculty jobs. In the second, participation in the contest is optional, as might be the case for established researchers who might compete to publish in the most prestigious journals but are not compelled to do so. 

The model illustrates how competition for limited rewards impels scientists to pursue higher-risk work in a world where other incentives often push in the opposite direction. We find that in mandatory contests, as competition intensifies (i.e., the ratio of prizes to participants drops) even risk-averse scientists will have to attempt riskier projects to have a chance of winning. Even when competition is slight in that almost everyone wins a prize, appreciable levels of risk-taking arise. In optional contests, as the stakes increase (i.e., the benefit of winning the contest increases relative to the value of opting out), increasing numbers of researchers elect to compete. This generates a negative externality for other scientists in the form of increasingly stringent competition and, critically, incentivizes additional risk-taking. As the size of the pool increases (i.e., the number of prizes and participants increases, holding their ratio constant), researchers become more homogeneous in the risk profiles that they adopt. The game analyzed here is a multi-prize extension of the classic and simpler ``silent duel'' game that was first studied in the post-war heyday of competitive game theory \cite{blackwell1954theory, karlin1959mathematical} and is now a standard example of a game of timing \cite{owen2013game}.  

\section{Model}

Our model considers a setting in which a group of scientists compete to win a limited number of prizes for generating the most valuable results. For simplicity, we assume that all prizes have equal value.  First, we consider the case where participation in the contest is mandatory, as might befit the setting of junior researchers competing for jobs.  Later, we consider a case where participation in the contest is voluntary, as might be the case when established researchers decide whether or not to try to publish in an elite journal. 

Researchers compete by making simultaneous one-time decisions about how risky of a project to pursue, in that each selects a research problem from some feasible region in risk-reward space. It makes sense to choose a project on the risk-reward frontier rather than inside it, but where on the frontier is optimal? How much risk should one be willing to adopt in exchange for potential reward?

Suppose that all scientific projects either succeed or fail and that projects can be indexed by the probability $p \in [0,1]$ that they succeed.  We can think of $p$ as a project's certainty and $1-p$ as the project's risk.  Projects that succeed generate results with scientific value (e.g., in terms of excitement, novelty, and impact) of $v(p)>0$. We assume the risk-reward frontier exhibits a trade-off such that successful risky projects provide greater scientific value than successful safe projects, i.e., $v'<0$.\footnote{We can be more explicit here.  Suppose that projects are characterized by the pair $(v, p)$. High-value, high-certainty projects are unavailable, either because they have already been claimed by past researchers, or because the attractiveness of the project will attract a crowd of other interested researchers, which in turn increases the risk of being scooped and thus left empty-handed \cite{kitcher1990}.  Thus, for any $p$, the available projects have values $v \in [0, v(p)]$, where $v(p)$ is the value of the highest-value project available for risk-level $p$.  Researchers intent on generating scientifically valuable outcomes would not choose a project with a value lower than $v(p)$ for a given $p$, and thus they only seriously contemplate projects with risk-level $p$ and value $v(p)$.  The fact that $v'<0$ reflects the trade-off between risk and reward along this frontier.}  Projects that fail are assumed to provide zero value.  Even the safest successful project generates more value than a failed project: $v(1) > 0$.

In game-theoretic language, this is a simultaneous-move game of complete information.  Researchers simultaneously choose an action $p \in [0,1]$, Nature determines whether the researchers' projects succeed or fail, and then prizes are given to the scientists with the most valuable outcomes.  Ties between researchers with equally valued outcomes are broken by drawing lots. If there are not enough successful projects to claim all the prizes, the unclaimed prizes go unawarded.\footnote{One might also consider a variation of the game in which unclaimed prizes are given out by chance to investigators with failed projects, if there aren't enough researchers with successful projects to claim all the prizes.  Whether or not any unclaimed prizes are given out does not impact our qualitative conclusions.} We use the Nash equilibrium (NE) as the solution concept and analyze for symmetric equilibria.  

As a historical note, the simplified setting in which two players compete for a single prize, and in which the prize is given to one of the two players at random if both players fail, is a classic problem from the early days of game theory known as the silent duel problem.\footnote{The moniker ``silent duel'' derives from the idea that pistol-wielding duelists of yore faced a tradeoff between how quickly they shoot and the accuracy of their shot.  The duel in this case is ``silent'' because neither player can hear (or otherwise detect) when the other fires their weapon.  Henig \& O'Neill \cite{henig1992games} call these games ``games of boldness'', which is a better fit for the context that we consider here.}  Its formulation is attributed chiefly to David Blackwell \cite{blackwell1954theory, radzik1996results}, and its Nash equilibrium, in which both players follow a mixed strategy on $p \in (1/3, 1]$, is discussed in detail in a variety of texts including e.g.\ \cite{karlin1959mathematical} and \cite{dresher1961mathematics}.  The extension to several players competing for a single prize has been solved both in its base version \cite{presman1977choice, sakaguchi1978marksmanship, henig1992games} and under various scenarios, including competitions among players with different skills \cite{presman2006existence}, generalized consolation prizes in the event that all fail \cite{alpern2019short}, and in the limit as the number of players becomes large \cite{hilhorst2018mixed}.  Our analysis below most directly parallels the path of Henig \& O'Neill \cite{henig1992games} and is, to our knowledge, the first to consider silent-duel-type games with several prizes or with voluntary participation.  Alpern \& Howard \cite{alpern2019short} have also noted the connection between silent-duel-type games and competition among scientific researchers.  

\section{Analysis}

\subsection{Obligatory participation}

Consider first a version of the game in which participation is obligatory and the number of players is finite.  In notation, suppose that $N \geq 2$ researchers compete for $K<N$ prizes.  It is easy to see that there cannot be a pure-strategy Nash equilibrium.  The intuition here is clear.  Consider the perspective of a focal researcher (dubbed ``Focal'').  If all of Focal's competitors play $p$, Focal's best response is to choose a project that is incrementally riskier than $p$ (say $p-\epsilon$, for some $\epsilon>0$). By doing so, Focal guarantees that they will receive a prize if their project succeeds---and avoids risking a tie with their competitors and settling for a prize awarded by lottery---and does so with only negligible additional scientific risk.  It follows that any NE must entail a mixed-strategy.  

As it turns out, this model falls within the scope of Dasgupta \& Maskin \cite{dasgupta1986existence}, whose results establish that a (symmetric) mixed-strategy Nash equilibrium (MSNE) exists and that it does not contain any atoms.\footnote{The appendix fills in the mathematical details to show that Dasgupta \& Maskin's results apply.} We now turn to characterizing the MSNE.  Let $f(p)$ and $F(p)$ denote the probability density function (pdf) and cumulative density function (cdf) of the MSNE, respectively.  Suppose that the MSNE has support on a subset $\mathcal{P} \subset [0, 1]$, and write $a = \inf \mathcal{P}$ and $b = \sup \mathcal{P}$.  To derive the payoff function at equilibrium, suppose Focal plays $p \in \mathcal{P}$.  Focal receives a prize if and only if their project succeeds and no more than than $K-1$ competitors succeed with projects that are riskier than Focal's project.  Let $\psi(p) = \int_a^p \, t \, f(t) \, dt$ be the probability that a competitor attempts a project risker than $p$ and is successful, and let $Y(p) \sim \mathsf{Binom}\bigl(N-1, \psi(p)\bigr)$ be the number of competitors who choose projects riskier than $p$ and obtain a successful outcome.  Let $\zeta(p) = \Pr{ Y(p) < K}$ be the probability that there are fewer than $K$ such competitors.  If we normalize the value of a prize to 1, then Focal's payoff to playing any $p \in \mathcal{P}$ is 
\begin{equation}
	\pi(p) = p \times \zeta(p).
	\label{eq:payoff-obligatory}
\end{equation}
This payoff must be the same for all $p \in \mathcal{P}$.   It readily follows that $\mathcal{P}$ must be a connected interval with supremum $b=1$, that is, $\mathcal{P} = (a, 1)$.\footnote{To show that $b=1$, suppose we have a MSNE with $b<1$.  Then the payoff to playing $b$ is $b \times \zeta(b)$.  Now consider playing $c \in (b, 1]$.  We must have $\zeta(b) = \zeta(c)$ (because no competitor will play an action $>b$).  Thus the payoff to playing $c$ is $c \times \zeta(c) = c \times \zeta (b) > b \times \zeta(b)$. This contradicts the assumption that we have a MSNE, so the MSNE must have $b=1$.  A nearly identical argument shows that the MSNE can't be a union of non-overlapping intervals.}  

The rest of the characterization is simply a numerical procedure that uses the properties of the binomial distribution to solve for the cdf of the MSNE.  Write the equilibrium payoff as $\pi^*$, and suppose $\pi^*$ is known. We can use $\pi^*$ to find $\zeta(p) = \pi^* / p$, and use $\zeta(p)$ to find $\psi(p)$.   To get $F(p)$ from $\psi(p)$, first differentiate $\psi(p)=\int_a^p \, t \, f(t) \, dt$ to give $\psi'(p) = p f(p)$ and hence $f(p) = \psi'(p) / p$.  Integrate to give $F(p)$:
\begin{displaymath}
	F(p)  = \int_a^p \, f(t) \, dt = \int_a^p \, \dfrac{\psi'(t)}{t} \, dt = \dfrac{\psi(p)}{p} + \int_a^p \, \dfrac{\psi(t)}{t^2} \, dt
\end{displaymath}
where the last equality uses a routine integration by parts. Finally, solve numerically for the value of $\pi^*$ that gives $F(1) = 1$.\footnote{Alpern \& Howard \cite{alpern2019short} give an exact solution when $K=1$, which can be used to verify the numerical solutions when $K=1$.}

\begin{figure*}[t]
	\begin{center}
		\includegraphics[width = 5in]{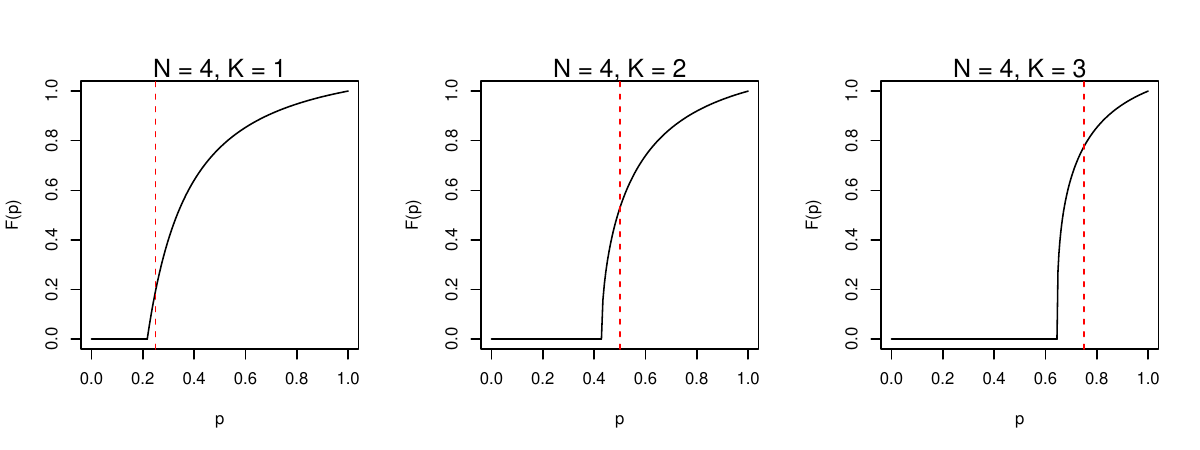}
		\caption{Cdfs of mixed-strategy Nash equilibria when $N=4$ investigators compete for $K=1$, 2, or 3 prizes and participation in the competition is obligatory.  Vertical red lines show the value $K/N$, which is the pure-strategy Nash equilibrium of the limiting case in which $N$ and $K$ grow large while maintaining the same proportion of prizes.}
		\label{fig:msne-vary-K}
	\end{center}
\end{figure*}

\begin{figure*}[t]
	\begin{center}
		\includegraphics[width = 5in]{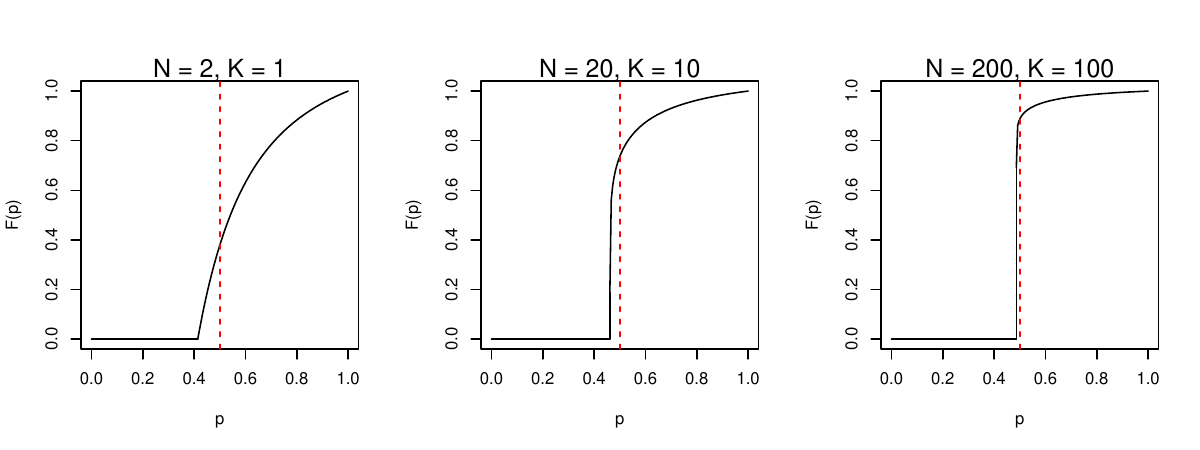}
		\caption{Cdfs of mixed-strategy Nash equilibria when $N=2$, 20, or 200 investigators compete for $K=1$, 10, or 100 prizes, respectively, and participation in the contest is obligatory.  Vertical red lines show the value $K/N$, which is the pure-strategy Nash equilibrium of the limiting case in which $N$ and $K$ grow large while maintaining the same proportion of prizes.}
		\label{fig:msne-vary-N}
	\end{center}
\end{figure*}

Figures~\ref{fig:msne-vary-K} and \ref{fig:msne-vary-N} illustrate the cdfs of the MSNEs for various combinations of $N$ and $K$.  Intuitively, decreasing the availability of prizes compels researchers to take larger scientific risks (Fig.~\ref{fig:msne-vary-K}).  Less obviously, if the proportion of available prizes ($K/N$) is held constant, then the MSNE predicts more variation in risk-taking behavior in small communities than in large ones (Fig.~\ref{fig:msne-vary-N}).  The intuition here is that in small communities there are two paths to winning a prize: one can attempt a risky project and hope to succeed, or attempt a safe project and hope that enough of the competition fails in their attempt at a risky project.  The spread of behaviors predicted in the MSNE for small communities reflects a blend of these two strategies.  In larger communities the law of large numbers makes it increasingly unlikely that the researchers playing the risk-taking strategy will leave enough prizes unclaimed to make the play-it-safe strategy pay off.  Hence, as the pool of competitors grows, the play-it-safe strategy becomes less viable, even if number of prizes grows in proportion to the number of competitors.

\begin{figure}[h!]
	\begin{center}
		\includegraphics[width = \linewidth]{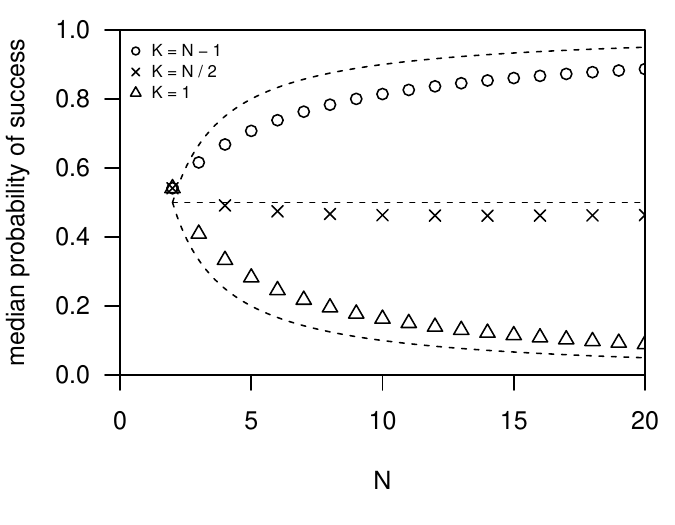}
		\caption{Median certainty, $p$, of the mixed-strategy Nash equilibrium (MSNE) when $N$ researchers compete for $K=N-1$ ($\circ$), $N/2$ ($\times$), or 1 ($\triangle$) prize(s).  Dashed lines show corresponding values of $K/N$, that is, the proportion of researchers who receive prizes if all prizes are awarded.}
		\label{fig:statics}
	\end{center}
\end{figure}

A second interesting observation is that even a small amount of competition induces more risk-taking than one might naively expect, at least in small-to-moderately sized communities.  Fig.~\ref{fig:statics} shows the median risk level of the MSNE for communities of size $N=2, \ldots, 20$ when there are enough prizes for all but one researcher, half of the researchers, or exactly one researcher.  Notice in particular that when the number of prizes is $K=N-1$, most researchers will select a project with certainty $p<K/N$.  (See also the rightmost panel in Fig.~\ref{fig:msne-vary-K}.)  Thus, even a little competition induces an appreciable increase in risk-seeking behavior.

The results above (Fig.~\ref{fig:msne-vary-N}) suggest that the MSNE becomes less diffuse as the number of competing scientists increases.  Indeed, this can be established directly by considering the limit as $N$ and $K$ go off to infinity with $K/N \rightarrow \phi \in (0, 1]$.  It is easy to see in this case that a pure-strategy Nash equilibrium is $p =\phi$.  Fig.~\ref{fig:msne-vary-N} suggests that the MSNE of the finite-$N$ game converges smoothly to the pure-strategy NE of the population game as $N$ and $K$ grow large.\footnote{See Hilhorst and Appert-Rolland \cite{hilhorst2018mixed} for a precise description of the rate of convergence when $K=1$ and the prize is awarded randomly if all fail.}

\subsection{Voluntary participation}

For all the similarities between the competition for jobs and the competition for elite publications, there is a critical difference: researchers can opt out of the latter. An up-and-coming researcher can't eschew the search for a faculty job if they want to stay in science.  However, more established researchers can opt out of the race for slots in glossy journals once they hold a tenured position. Here, we explore how our model predictions change when researchers have the option not to participate in the competition at all.

Suppose that researchers looking to publish their work have two options.  They can either try to publish their work in an elite journal, or they can publish their work in a less-competitive, discipline-specific journal.  Researchers obtain a reward normalized to 1 if they publish in the discipline-specific journal, and they obtain a reward $1 + \beta$ if they publish in the elite journal, where $\beta>0$ is the additive premium for publishing in the high-profile venue. Both journals only publish successful studies; the glossy journal publishes the most valuable $K<N$ successful studies (and publishes fewer than $K$ studies if fewer than $K$ studies succeed).  Finally, we assume that a researcher who generates a successful study but is rejected by the glossy journal  can still publish in the discipline-specific journal for the full reward of 1.  This model with voluntary participation approaches the model with obligatory participation as the premium $\beta$ becomes large.

The equilibrium of a model with voluntary participation can be characterized using nearly an identical approach to the previous model.  Using the same definitions as before---in particular, continuing to define $\zeta(p)$ as the probability that fewer than $K$ competitors succeed with projects riskier than $p$---the payoff function to playing $p$ under voluntary participation becomes 
\begin{equation}
	\pi(p) = p \left[ 1 + \beta \zeta(p) \right]
	\label{eq:payoff-voluntary}
\end{equation}
(compare eq.~\ref{eq:payoff-voluntary} with eq.~\ref{eq:payoff-obligatory}).  Equipped with this payoff function, the same solution strategy as before can be used to find the cdf of the MSNE numerically.

\begin{figure*}[t]
	\begin{center}
		\includegraphics[width = 6in]{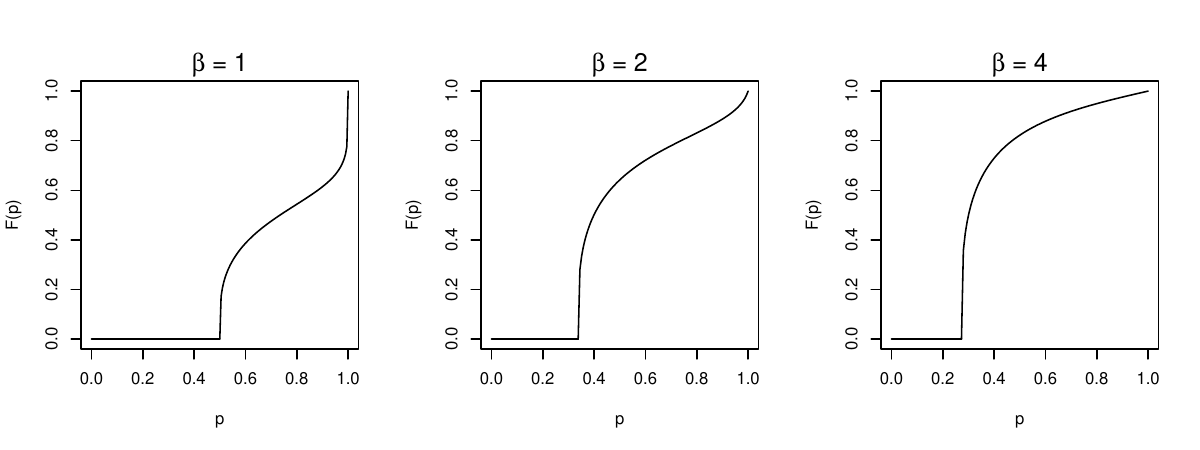}
		\caption{Cdfs of mixed-strategy Nash equilibria when $N=20$ researchers compete for $K=5$ slots in a glossy journal, and participating in the competition is voluntary.  Panels show cases in which the additive premium for publishing in a glossy journal is $\beta = 1$, $\beta=2$, or $\beta = 4$.  Nash equilibria in the equivalent large-$N$, large-$K$ game with $\phi=0.25$ (not shown) are as follows. When $\beta=1$, half play $p=0.5$, and half play $p=1$. When $\beta = 2$, three-quarters play $p=1/3$ and one-quarter play $p=1$. When $\beta = 4$, all play $p=0.25$.}
		\label{fig:msne-backup-journal}
	\end{center}
\end{figure*}

Fig.~\ref{fig:msne-backup-journal} shows the MSNE when $N=20$ researchers compete for $K=5$ prizes under several values of $\beta$.  When the premium $\beta$ is large, the MSNE resembles the MSNE of the obligatory-participation game.  However, when the premium is small, the MSNE blends two distinct strategies: a fraction of researchers try for the glossy journal, while a second fraction of researchers play it safe and play $p \approx 1$.   The fact that some researchers opt out of the competition for the glossy journal reduces the competition among the researchers who do try for the glossy journal, allowing those researchers to take less of a scientific risk.  

\begin{figure}[h!]
	\begin{center}
		\includegraphics[width = 3in]{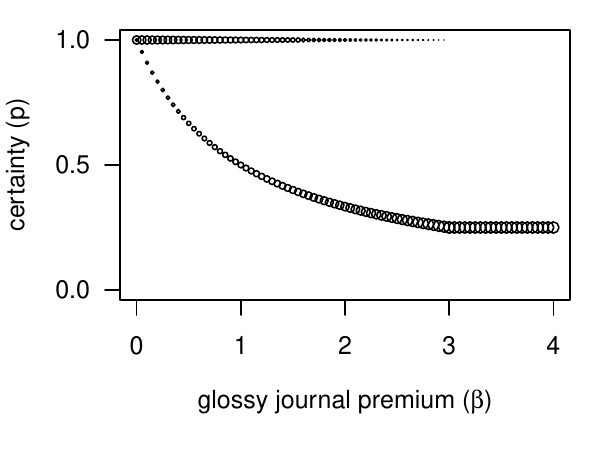}
		\caption{Nash equilibria when a large number of researchers compete for slots in a glossy journal, and participating in the competition is optional in the sense that a discipline-specific journal will publish any successful study.  $\beta$ gives the additive premium for publishing in the glossy journal. The glossy journal can publish at most $\phi=25\%$ of the researchers' studies.  For $\beta<3$, the Nash equilibrium is a two-part mixture.  Size of the plotting symbols corresponds to the proportion of the population that plays each component of the equilibrium.  The Nash equilibrium is a pure equilibrium when $\beta\geq3$.}
		\label{fig:msne-backup-population}
	\end{center}
\end{figure}

For the voluntary-participation game, the large-$N$ limit is especially clarifying.  Suppose a unit mass of investigators compete and the glossy journal can publish at most a fraction $\phi$ of the researchers' studies.  If $\phi(1 + \beta) \geq 1$, all investigators play the pure-strategy NE $p=\phi$, just like the obligatory-participation model.  However, if $\phi(1 + \beta) < 1$, then the NE is a two-part mixture, in which a fraction $\phi(1+\beta)$ of the researchers aim for the glossy journal and play $p=1/(1+\beta)$, while the remainder of the researchers play it safe and play $p=1$ (Fig.~\ref{fig:msne-backup-population}).  Thus, the premium that the community attaches to publishing in the glossy journal has substantial impacts across a number of dimensions. First, it impacts willingness of researchers to employ the play-it-safe strategy. Second, it determines the intensity of competition among researchers who try to publish in the glossy journal. Third, as the intensity of competition increases so does risk-taking, and thus the premium associated with the elite journal shapes the risk behavior of researchers in the field.\footnote{Note also that as $\beta$ increases fewer successful projects are available to be published in the discipline-specific journal.  Moreover, once $\beta$ becomes large enough, the discipline-specific journal publishes nothing but papers rejected from the glossy journal.} 

\subsection{Heterogeneous research ability}

We can also explore how play changes when researchers differ in their abilities to generate successful results.  Karlin \cite{karlin1959mathematical} and Dresher \cite{dresher1961mathematics} provide solutions when players of different abilities compete in the ``silent duel'' game.\footnote{Recall that in the silent-duel game participation is obligatory and the prize is given randomly to one of the two researchers if both fail.  The awarding of the prize randomly when both players fail shifts the researchers towards riskier play compared to our primary model in which some prize(s) goes unawarded if fewer than $K$ researchers' projects succeed.  However, for the present purpose, the qualitative effect of differences in researcher abilities on equilibrium play will be the same regardless of whether or not the prize is awarded if both researchers fail.  Thus we borrow the solutions in \cite{karlin1959mathematical} and \cite{dresher1961mathematics} here.} That work provides us with intuition about how play changes in our game when researchers' abilities differ. 

\begin{figure}[h!]
\begin{center}
	\includegraphics[width = 3in]{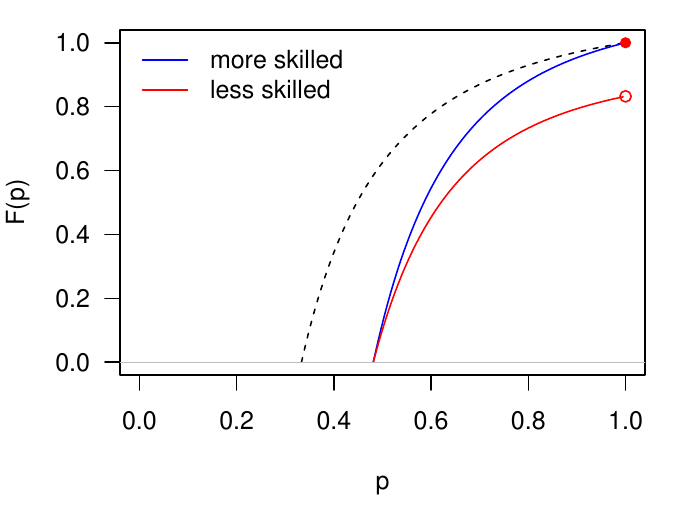}
	\caption{Cdfs of mixed-strategy Nash equilibria when $N=2$ researchers compete for $K=1$ prize, participation is obligatory, the prize is awarded randomly if both fail, and the players have different abilities.  The blue and red lines show the cdf of the MSNE for the more- and less-skilled researcher, respectively.  The less-skilled researcher's MSNE has a point mass (indicated by the discontinuity in the cdf) at $p=1$. The dashed black line shows the cdf of the MSNE when both researchers are equally skilled, and their skill is equal to that of the more-skilled researcher.}
	\label{fig:different-abilities}
\end{center}
\end{figure}

Suppose two researchers compete for a single prize, and let $p$ give the probability that the more skilled researcher obtains a successful outcome.  Suppose that, when attempting a project for which the more-skilled researcher succeeds with probability $p$, the less-skilled researcher succeeds with probability $p^2$.  Fig.~\ref{fig:different-abilities}, obtained using the solution given in citations \cite{karlin1959mathematical} and \cite{dresher1961mathematics}, shows that the more-skilled researcher attempts a stochastically riskier project than the less-skilled researcher.  Neither researcher is as risk-seeking as they would be if both were as skilled as the more-skilled researcher.  These results are largely intuitive and suggest that risk-seeking increases as researchers become more skilled and as their competition becomes more skilled.  We expect similar results would prevail in the more general case of several researchers competing for several prizes, although formal mathematical confirmation awaits future work.

\subsection{Training and collaboration}

In many areas of the natural sciences, training practices and funding structures result in a hierarchical organization of research effort, wherein senior investigators supervise teams of graduate students and postdoctoral researchers \cite{bozeman2004scientists,wuchty2007increasing,zhang2022labor}. Co-authorship is often the norm for these collaborations \cite{adams2005scientific,thelwall2022research}, and as a result the professional success of team members at disparate career stages becomes coupled.

Yet investigators at different career stages may be competing in very different contests with different prize thresholds \cite{fitzenberger2014up,fox2001careers}. A postdoc facing a tight job market may be competing in a contest with far more players than prizes, whereas an an assistant professor seeking tenure at a mid-tier institution may be participating in a contest where the vast majority of players will win. As a result, the incentives facing collaborators may differ considerably by career stage, resulting in considerable tension within a research team as to desirable levels of research risk. 

\begin{figure}[h!]
\begin{center}
	\includegraphics[width = \linewidth]{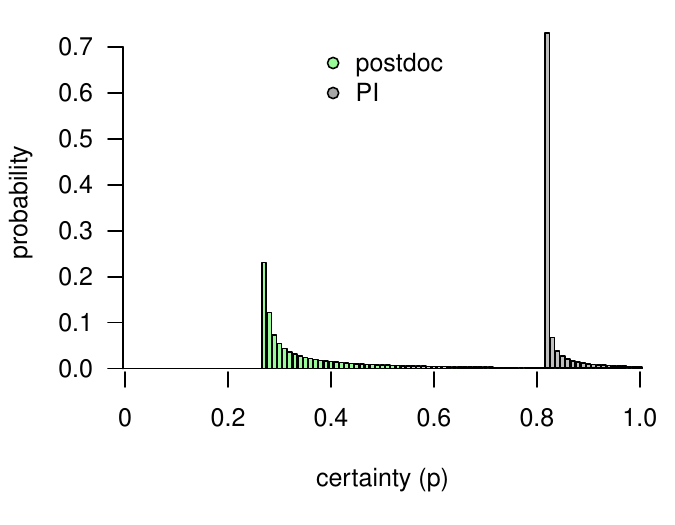}
	\caption{MSNEs for a postdoc (green) engaged in a competition among $N=10$ postdocs for $K=3$ jobs and an assistant professor (gray) engaged in a competition among $N=10$ assistant professors for $K=9$ promotions.  For ease of visual comparison, MSNEs are shown by the probability that each assigns to a certainly value $p$ in the intervals $[0, 0.01], (0.01, 0.02], (0.02, 0.03], \ldots, (0.99, 1]$. The postdoc's MSNE puts positive probability on all values of $p \geq 0.273...$, but the probability is too small to see for large values of $p$.}
	\label{fig:pdvspi}
\end{center}
\end{figure}

Figure \ref{fig:pdvspi} shows the MSNEs across risk levels for the two individuals mentioned above: a postdoc engaged in a competition among $N=10$ postdocs for $K=3$ jobs and an assistant professor engaged in a competition among $N=10$ assistant professors for $K=9$ promotions.. Notice that the MSNE for the assistant professor puts zero probability mass on much of the risk domain in support of the postdoc's mixed strategy. 

When competition is optional, the problem may be exacerbated as some team members wish to compete and others wish to opt out entirely. A graduate student looking for a prime postdoctoral position may want to compete to publish in an elite journal, whereas the student's tenured advisor may not want any part of that rat-race. 

Team formation in science and industry is often viewed in terms of complementarity of human capital, social capital, and financial capital \cite{becker1992division,ethiraj2012division,semrau2016complementary,ployhart2017human,neffke2019value}. Our model suggests that aligned incentives will be another important dimension of team success---and along this axis, successful matching favors similarity rather than complementary. 

\section*{Discussion}

Our model highlights an awkward bind in which academic scientists find themselves. Researchers are likely to be risk-averse in wages\footnote{Here we mean wages broadly construed to include all of the rewards, monetary or otherwise, that scientists value: salary, job security, prestige, research funding, independence, etc.}. Since academic researchers are rewarded for their output rather than their effort \cite{dasgupta1987information}, as risk-averse agents we expect them to choose low-risk projects even if this reduces the expected value of their discoveries \cite{gross2024rationalizing}. The contests we model here push in the opposite direction, because while output is rewarded rather than effort, that output is only rewarded if clears some bar in a competitive interaction. To clear this bar, researchers are impelled to choose risky projects. They are also forced to now internalize two forms of risk---first, the risk of project failure from the riskier projects they choose, and second, the risk of being out-competed and failing to receive a reward for even a successful project. 

On its face,  a mixed-strategy Nash equilibrium may seem to be an implausible description of how scientists choose projects. Surely, researchers do not flip coins when selecting a research direction on which to embark.  This is a common criticism of mixed-strategy Nash equilibrium in games of complete information \cite{fudenberg1991game, osborne1994course}.  We can justify the MSNE here by appealing to Harsanyi's purification theorem \cite{harsanyi1973games}. This theorem states that MSNEs of games of complete information can be regarded as limits of pure-strategy equilibria of nearby games of incomplete information.  In those nearby games, each player's payoff function is independently perturbed by a small amount.  Small, idiosyncratic differences among researchers' payoff functions can arise in a variety of ways, such as researchers deriving varying degrees of intrinsic satisfaction from making a discovery that creates utility above and beyond winning the prize.  The resulting asymmetric game would give rise to a pure-strategy NE\footnote{We assume that variation in payoff functions is continuously distributed, so that the odds of two researchers having identical payoff functions in the perturbed game is nil.} in which each player's equilibrium strategy depends on the particular payoff function that face.  Now, consider the distribution of pure-strategy NE that each player might choose {\em ex ante} before they learn their particular payoff function, and take the limit of these distributions as the differences among researchers' payoff functions shrink to zero.  Harsanyi's result establishes that the limit will coincide with the MSNE of the symmetric game, regardless of how the idiosyncrasies in researchers' payoffs arise.\footnote{Harsanyi's result only applies to games with finite action spaces.  Thus, to be rigorous, we would have to apply Harsanyi's result to a discretization of our game in which players can only choose actions on a discrete set of values for $p$ in $[0,1]$.  However, there is no limit on how fine this discretization of the action space can be, and in reality researchers cannot choose the riskiness of their project to infinite precision anyhow.  So we have no qualms arguing that Harsanyi's result should apply with full force to our setting as well.  A simple illustration of Harsanyi's purification argument in our model appears in the appendix.}  Thus, the MSNE can be viewed, and perhaps is more satisfyingly viewed, as the {\em ex ante} distribution of strategies that arise in equilibrium when researchers' payoff functions vary in small and idiosyncratic ways, as opposed to a prediction of genuine randomization by researchers in a perfectly symmetric setting.

One might also wonder how the assumption of perfect discrimination by the judges of the contest (hiring committees or journal editors) affects our predictions.  Surely, the scientific value of a completed project is not a quantity that can be objectively measured to arbitrary precision.  It would be straightforward to introduce evaluation noise into our model by layering on a random assessment score of the value of a completed project.  It is clear to see that as the evaluation becomes increasingly noisy, researchers have less of an incentive to pursue risky projects, as long as the distinction between a ``successful'' and ``unsuccessful'' project remains clear.\footnote{Which it would if, for example, successful projects are published and unsuccessful projects are not.}  Indeed, in the limit as the assessment of scientific value becomes infinitely noisy, there is no incentive at all to take on any scientific risk, and the NE under any scenario is for every researcher to pursue the most certain possible project.  Intermediate levels of noise in evaluation lead to equilibria that intergrade smoothly between this case and the MSNE under perfect evaluation.

Our models highlight the interconnectedness of the processes that shape scientific incentives and thus influence scholars' research strategies. Changes in the intensity of competition for various scientific positions and accolades do not merely have screening effects on the labor force; they end up restructuring the type of work that is conducted. Risk profiles shift, and with those shifts come cultural changes in what is considered valuable work---particularly for those researchers at career stages where they cannot choose to opt out of the race.

\section*{Acknowledgments}

This work was partially supported by NSF awards SES-2346645 to CTB and SES-2346644 to KG, and by TWCF Diverse Intelligences frameworks grant 32581 to CTB.  We thank Kevin Zollman for helpful feedback, and KG thanks P\'eter Bayer for early stimulating discussions and the University of Washington Department of Biology for visitor support.
	
	\section*{LLM use statement}
	We used ChatGPT 5 to spot-check numerical results.  All of the text and mathematical results are our own human outputs.
	
	\bibliography{game}
	
	\newpage 
	
	\renewcommand\theequation{A.\arabic{equation}}  
	\setcounter{equation}{0}   
	
	\renewcommand{\thefigure}{A.\arabic{figure}}
	\setcounter{figure}{0}
	
	\renewcommand{\thesubsection}{A.\arabic{subsection}}
	\renewcommand{\thesubsubsection}{A.\arabic{subsection}.\arabic{subsubsection}}
	\setcounter{subsection}{0}  
		
	\onecolumngrid
	\section*{Appendix}

	\subsection{Proof of the existence of a MSNE}
	
	That a MSNE exists in our game follows from Theorem 5 of Dasgupta \& Maskin \cite{dasgupta1986existence} (henceforth DM86).  Here, we fill in the details to confirm that the proof applies for the obligatory-participation game.  A nearly identical argument applies to the voluntary-participation game.  Helpfully, DM86 illustrate their results with the silent-duel game, and their application of the pertinent theorems to the silent-duel game also applies to our game with only trivial changes.
	
	Consider the original version of the game in which Nature is not a player.  In the usual way, write $\mathbf{p}_{-i} = (p_1, \ldots, p_{i-1}, p_{i+1}, \ldots, p_N)$ as the profile of actions of all researchers except researcher $i$. Theorem 5 of DM86 shows that a MSNE for our game will exist if three non-trivial conditions hold. First, the payoff functions $\pi_i(\mathbf{p})$ must be continuous except on a sub-manifold of the action space of Lebesgue measure zero.  In our game, the payoff functions are continuous everywhere in the hypercube $[0,1]^N$ except on the diagonals where $p_i=p_j$ for some $i\neq j$ (excluding the origin).  These diagonals form such a sub-manifold as the theorem requires.
	
	Second, the sum of the payoffs $\sum_{i=1}^N \pi_i(\mathbf{p})$ has to be upper semi-continuous in each $p_i$.  But for our game $\sum_{i=1}^N \pi_i(\mathbf{p}) = \min(K, \sum_{i=1}^N p_i)$, which of course is continuous in $p_i$ and hence upper semi-continuous also. 
	
	Third, Theorem 5 of DM86 requires that $\pi_i(p_i, \mathbf{p}_{-i})$ is weakly lower semi-continuous in $p_i$. Following Definition 6 of DM86, a sufficient condition for weak lower semi-continuity is left lower semi-continuity, that is, for any point $\mathbf{p}$ at which $\pi_i(p_i, \mathbf{p}_{-i})$ is discontinuous and for any sequence $\left\{p^n_i\right\}$ that converges to $p_i$ from below, $\lim_{p^n_i \uparrow p_i} \pi_i(p^n_i, \mathbf{p}_{-i}) \geq \pi_i(p_i, \mathbf{p}_{-i})$.  (Take $\lambda = 1$ in Definition 6 of DM86.) But discontinuities occur precisely when researchers choose exactly the same action as a competitor, in which case they are at risk of having to draw lots for the prize if both the researcher's project and their competitor's project succeeds. This drawing of lots will lower the researcher's payoff discontinuously, thus satisfying the needed condition.\footnote{Note that each researcher's payoff is continuous at $p=0$ (their payoff is 0 regardless of what their competitors do), so we do not have to consider the impossibility of approaching $p=0$ from below.} 
	
	That the MSNE is atomless follows from Theorem 6 of DM86.  Again, DM86's application to the silent-duel game applies to our game essentially unchanged.
	
	\subsection{A purification illustration}
	
	Here we illustrate that Harsanyi's purification argument \cite{harsanyi1973games} holds for a simplified version of our model.  That Harsanyi's argument holds in this case is not a surprise, of course; Harsanyi's results \cite{harsanyi1973games, govindan2003short} guarantee it.  Nonetheless, an illustration edifies.
	
	Consider a simplified version of the obligatory-participation model in which $N=2$ researchers compete for $K=1$ prize.  In this simplification, each researchers' action space is binary: either they can choose a risky project ($R$) that has a probability $p$ of succeeding, or they can choose a safe project ($S$) that is certain to succeed.  The probability $p$ is given exogeneously.  The rest of the game proceeds as before, with no prize awarded if both fail.  
	
	Formally, let $\mathcal{A}_i = \left\{R, S\right\}$ be the action space for player $i=1,2$.  The playoff function for player 1 is
	\begin{displaymath}
		\pi_1(a_1, a_2) = \begin{cases}
			p ( 1 - p / 2) & a_1 = a_2 = R \\
			p & a_1 = R, a_2 = S \\
			1-p & a_1 = S, a_2= R \\
			1/2 & a_1 = a_2 = S.
		\end{cases}
	\end{displaymath}
	Player 2's payoff function is similar.
	
	It is easy to show that if $p \in (1/2, 2 - \sqrt{2})$ then the game has a symmetric MSNE in which $S$ is played with probability $q = 1 - (2p-1) / (p^2 - 2p + 1) = (p^2-4 p+2)/(1-p)^2$
	
	We now use Harsanyi's argument to show that the MSNE can be interpreted as the limit of a Bayesian Nash equilibrium (BNE) in a nearby perturbed game of incomplete information.  Suppose now that if player $i$ plays $R$, then their probability of success is $p + \epsilon\, t_i$, where $t_i \sim \mathsf{Unif}(0,1)$, $t_1$ and $t_2$ are independent, and $\epsilon$ is small, such that $p + \epsilon < 1$.  We refer to $t_i$ as player $i$'s type. This game has a symmetric BNE in which each player plays $S$ iff $t < c$ for some threshold $c \in [0,1]$, and each player believes (correctly) that the other player will play $S$ with probability $c$.  In the usual way, we assume that the timing of the game is: (1) players learn their own type, (2) players choose an action, and (3) payoffs are realized.
	
	For the moment, allow for the possibility that $c$ may differ between the players, and write $c_i$ as the threshold type for player $i$.  Of course, in a symmetric equilibrium we will have $c_1 = c_2$.
	
	It is straightforward to derive payoff functions for each player.  Consider player 1.  Upon learning their type, player 1's payoff to playing $R$ is
	\begin{displaymath}
		\pi_1(R, t_1) = (p + \epsilon\, t_1) \left[ 1 - \frac{1}{2}(1 - c_2)\left(p + \frac{\epsilon(1 + c_2)}{2}\right)\right].
	\end{displaymath}
	In the expressison above, the term $(1 - c_2)\left(p + \frac{\epsilon(1 + c_2)}{2}\right)$ gives the probability that player 2 plays $R$ and succeeds.  In that expression, the term $1 - c_2$ is the probability that player 2 plays $R$. (Or, more precisely, it is player 1's correct belief of this probability.)  The term $p + \frac{\epsilon(1 + c_2)}{2}$ is player 2's chance of success if they play $R$, where $ \frac{1 + c_2}{2}$ is player 2's expected type given that they play $R$.
	
	Player 1's payoff to playing $S$ is
	\begin{displaymath}
		\pi_1(S, t_1) = (1 - c_2)\left( 1 - \left(p + \frac{\epsilon(1 + c_2)}{2}\right)\right) + \frac{c_2}{2}.
	\end{displaymath}
	In the expression above, the first term on the right is the probability that player 2 plays $R$ and fails.  The term $c_2 / 2$ is the probability that player 2 plays $S$ and thus the prize is awarded by drawing lots.
	
	It is clear that $\pi_1(R, t_1)$ increases with $t_1$ while $\pi_1(S, t_1)$ does not depend on $t_1$.  Therefore player 1 will play $S$ iff $t_1 < c_1$ for some threshold $c_1$.  This threshold is the value that sets $\pi_1(R, c_1) = \pi_1(S, c_1)$.  Then, because the equilibrium is symmetric, we must have $c_1 = c_2 = c$.  Setting $\pi_1(R, c_1) = \pi_1(S, c_1)$ and $c_1 = c_2$ yields
	\begin{equation}
		(p + \epsilon c) \left[ 1 - \frac{1}{2}(1 - c)\left(p + \frac{\epsilon(1 + c)}{2}\right)\right] = (1 - c)\left( 1 - \left(p + \frac{\epsilon(1 + c)}{2}\right)\right) + \frac{c}{2}. 
		\label{eq:purification}
	\end{equation}
	Because we are interested in the limit as $\epsilon \rightarrow 0$, we can ignore the $\epsilon^2$ terms and this becomes a quadratic in $c$. Dropping those $\epsilon^2$ terms and expanding:
	\begin{equation}\frac{3}{4} c^2 \,p\, \epsilon +\frac{c \,p^2}{2}-\frac{c \,p \,\epsilon }{2}+c\, \epsilon -\frac{p^2}{2}-\frac{p \,\epsilon }{4}+p=c \,p-\frac{c}{2}-p-\frac{\epsilon }{2}+1
	\end{equation}
	And collecting the remaining terms:
	\begin{equation}3\, c^2 \,p\, \epsilon + 2 \,c \left(p^2-p (\epsilon +2)+2 \,\epsilon +1\right)+ \left(-2\, p^2-p\, \epsilon +8\, p+2 \,\epsilon -4\right)=0
	\end{equation}
	Applying the quadratic formula:
	\begin{equation}\frac{-2 \left(p^2-p (\epsilon +2)+2 \epsilon +1\right)\pm \sqrt{4 \left(p^2-p (\epsilon +2)+2 \epsilon +1\right)^2-12 \,p \,\epsilon  \left(-2 p^2-p \,\epsilon +8 p+2 \epsilon -4\right)}}{6\, p\, \epsilon }
	\end{equation}
	We now apply L'H\^opital's rule. The ratio of derivatives of numerator and denominator with respect to $\epsilon$ is
	\begin{equation}
		\frac{-2 (2-p)+\frac{8 (2-p) \left(p^2-2 p+1\right)-12 p \left(-2 p^2+8 p-4\right)}{4 \sqrt{\left(p^2-2 p+1\right)^2}}}{6 p}
	\end{equation}
	Simplifying, we get 
	\begin{equation}\frac{p^2-4 p+2}{(1-p)^2}\end{equation}
	which of course is the MSNE for the game. 
	
\end{document}